\documentclass[usenatbib]{mn2e}
\usepackage{graphicx}
\footnotesize
\newdimen\minuswidth    
\setbox0=\hbox{$-$}
\minuswidth=\wd0
\catcode`@=\active
\def@{\kern\minuswidth}
\newdimen\digitwidth    
\setbox0=\hbox{\rm0}
\digitwidth=\wd0
\catcode`!=\active
\def!{\kern\digitwidth}
\normalsize
\title[Are the infrared-faint radio sources pulsars?]
{Are the infrared-faint radio sources pulsars?}
\author[A. D. Cameron et al.]
{
A. D. Cameron,$^{1,2}$
M. Keith,$^2$
G. Hobbs,$^2$
R. P. Norris,$^2$
M. Y. Mao$^{2,3,4}$ \& 
\newauthor
E. Middelberg$^5$
\\
$^1$ School of Physics, University of New South Wales, Sydney, NSW 2052, Australia\\
$^2$ CSIRO Astronomy \& Space Science, PO~Box~76, Epping
NSW~1710, Australia \\
$^3$ School of Mathematics and Physics, University of Tasmania, Private Bag 37, Hobart, 7001, Australia\\
$^4$ Anglo-Australian Observatory, PO Box 296, Epping, NSW, 1710, Australia\\ 
$^5$ Astronomisches Institut, Ruhr-Universit\"at Bochum, Universit\"atsstr. 150, 44801 Bochum, Germany \\
}

\date{}

\begin{document}
\maketitle
\newcommand{\setthebls}{
}
\setthebls
\begin{abstract}

Infrared-Faint Radio Sources (IFRS) are objects which are strong at radio wavelengths but undetected in sensitive Spitzer observations at infrared wavelengths. Their nature is uncertain and most have not yet been associated with any known astrophysical object. One possibility is that they are radio pulsars. To test this hypothesis we undertook observations of 16 of these sources with the Parkes Radio Telescope. Our results limit the radio emission to a pulsed flux density of less than 0.21 mJy (assuming a 50\% duty cycle). This is well below the flux density of the IFRS. We therefore conclude that these IFRS are not radio pulsars.

\end{abstract}


\section{Introduction}\label{sec:Intro}
Infrared-Faint Radio Sources (IFRS) are objects which typically have flux densities of several mJy at 1.4 GHz, but are undetected at 3.6 $\mu$m using sensitive \emph{Spitzer Space Telescope} observations with $\mu$Jy sensitivities \citep{naa+06}. They were an unexpected discovery in the Australia Telescope Large Area Survey (ATLAS), which, covering seven square degrees at 1400 MHz to $\sim$30 $\mu$Jy, is  the widest deep field radio survey attempted thus far (Norris et al. 2006; Middelberg et al. 2008a)\nocite{naa+06,mnc+08}. There are $\sim 50$ IFRS in ATLAS, accounting for 2.5\% of the current ATLAS source catalogue.  Most of the IFRS are unresolved at 1.4\,GHz and have flux densities of up to tens of mJy. 

The ATLAS project surveys two regions of sky. The first region includes the Chandra Deep Field South (CDFS) \citep{grt+01} which encompasses the Great Observatories Origins Deep Survey (GOODS) field \citep{gfk+04}.  The second region coincides with the European Large Area ISO Survey-South1 (ELAIS-S1). Both regions have also been covered by the \emph{Spitzer} Wide-Area Infrared Extragalactic (SWIRE) survey program \citep{lsrr+03}.  The second region is close to (but does not overlap) the region covered by the recent Parkes High Latitude Pulsar Survey \citep{bjd+06}. 

The primary goals of ATLAS are to trace the cosmic evolution of active galactic nuclei (AGN) and star-forming galaxies. The detected sources are expected to follow relatively well-known spectral energy distributions (SED) that predict infrared emission detectable by \emph{Spitzer} from most strong radio sources. 
Since the discovery of IFRS in 2006, several publications have attempted to understand their nature and emission mechanisms. Very long baseline interferometry (VLBI) observations of a total of six IFRS by \citet{Norris07b} and Middelberg et al. (2008b)\nocite{mnt+08} have resulted in the detection of high-brightness temperature cores in two IFRS. Recently, \citet{hms+10} investigated the four IFRS in the GOODS field, for which ultra-deep Spitzer imaging had recently become available, and attempted to fit template SEDs corresponding to known classes of galaxies and quasars. Norris et al. (2010)\nocite{norris2010} have extended this work using deep Spitzer imaging data from the Spitzer Extragalactic Representative Volume Survey project (Lacy et al. 2010\nocite{lacy2010}), and have shown that most IFRS sources have extreme values of S20/S3.6, which can be fitted by a high-redshift  radio-loud galaxy or quasar. They also stacked the deep Spitzer data to show that a typical IFRS must have a median 3.6 $\mu$m flux density of no more than 0.2 $\mu$Jy, giving extreme values of the radio-infrared flux density ratio. Middelberg et al. (2011)\nocite{middelberg2010} have found the spectra of IFRS are remarkably steep, with a median spectral index of $-$1.4 and a prominent lack of spectral indices flatter than $-$0.7.  We note that the systematic steepening of spectral indices can not be the result of variability.

While unidentified radio sources have been well-known for many decades, they are usually identified once sufficiently sensitive optical/IR observations become available. The IFRS differ from previous classes of unidentified sources in their extreme ratio of radio/IR flux density. For example, the ratio of 20 cm to 3.6 $\mu$m flux density of the IFRS studied here is typically 1000, and is 8000 in the case of ES247 (Norris et al. 2010\nocite{norris2010}). This may be compared with the ratio of $~$10 expected for starburst galaxies and $~100$ for a typical radio-loud quasar template \citep{ewmd+94}. They bear some similarities to the high-redshift radio galaxies studied by \citet{ssd+07} and \citet{jts+09} but appear in many cases to be even more extreme. Hence, the nature of IFRS remains unclear, largely because they are invisible to even the most sensitive optical/infrared observations. While at least some are likely to be high-redshift radio galaxies or quasars, it is also likely that the class may encompass several types of object. Other candidates for IFRS include
heavily obscured AGN, stray radio lobes, radio relics and radio pulsars (\citealp{naa+06, Norris07b, ga08}; Middelberg et al. 2008b\nocite{mnt+08}). \citet{Norris07b} propose that IFRS may even represent a phase of AGN evolution.


For 16 of these sources the measured flux densities and spectral indices are consistent with those measured for pulsars in our own galaxy \citep{mhth05}\footnote{http://www.atnf.csiro.au/research/pulsar/psrcat/} .  As the ATLAS survey regions are well away from the Galactic plane, it is unlikely that all 16 of these sources are pulsars, but the possibility remains that some are.  If so, then it is important to identify them so that (a) they can be removed from the extragalactic source statistics, and (b) so they can be accounted for in models of pulsar populations.

However, it is difficult to estimate how many pulsars would be expected in any given region of the sky. The most recent pulsar surveys have covered regions close to the Galactic plane (e.g., \citealp{mlc+01}). These pulsar surveys have discovered a large number of new pulsars, but relatively few millisecond pulsars. The Galaxy is likely to contain a similar number of normal and millisecond pulsars \citep{lml+98}, but the millisecond pulsars are likely to have travelled significant distances from the Galactic plane and are difficult to find due to their relatively low luminosities, their fast spin-rates and because they are commonly in binary systems, which disrupts the regular periodicity of their emitted pulses. The only deep 20-cm pulsar survey covering a large area that is far from the Galactic plane is the Parkes High Latitude Pulsar Survey, which covers Galactic longitudes $220^\circ < l < 260^\circ$ and latitudes $|b|<60^\circ$ \citep{bjd+06}, with 6456 pointings of 265 seconds each. This survey detected 42 pulsars with a survey sensitivity of pulsed flux density $\sim0.5$ mJy. Even though this survey made relatively few discoveries, the pulsars detected away from the Galactic plane were of great astrophysical interest. They included the first double pulsar system PSR~J0737$-$3039 \citep{lbk+04} and three other millisecond pulsars. \citet{bjd+06} discussed the Galactic latitude distribution of their detections in detail, and suggested a roughly uniform distribution. However, this survey was relatively insensitive to millisecond pulsars, because of the restricted sampling time and number of frequency channels, and so the total number of millisecond pulsars detected was small. Hence, it is currently impossible to make a reasonable estimate of the number of pulsars likely to be detected per square degree at high Galactic latitude. With a survey sensitive to millisecond pulsars it seems reasonable that at least one detectable pulsar exists in the ATLAS survey region of roughly seven square degrees, but because this number is so poorly known, it adds a further motivation for this study.


It should be noted that detections and discoveries of pulsars in radio continuum data are not without precedent. Examples include PSR~B1937$+$21, the first millisecond pulsar \citep{bkh+82}, PSR~B1821$-$24, found during a VLA imaging search of globular clusters \citep{hhb+85}, PSR~B1951$+$32, initially identified as a steep-spectrum polarised point source central to the supernova remnant CTB 80 \citep{Strom+87}, and PSR~J0218$+$4232, originally uncovered by the Westerbork Synthesis Radio Telescope (WSRT) \citep{ndf+95}.

\section{Observations}

\nocite{middelberg2010}\nocite{mnc+08}

\begin{table*}
\caption{Parameters of the infrared-faint radio sources selected for analysis as potential pulsars. All flux density and spectral index data is obtained from Middelberg et al. (2011), with the exception of ES1259 (Middelberg et al. 2008a) and CS255 \citep{naa+06}.}\label{tb:sources}
\begin{tabular}{lllll}\hline
Source & Right ascension & Declination & 20cm Flux density & Spectral index \\
              & (hms)                    & (dms)           & (mJy)             & \\ \hline
ES011 & 00:32:7.444 & $-$44:39:57.8 &  	9.5 & $-$1.44 \\
ES318 & 00:37:05.54 & $-$44:07:33.7 & 2.0  & $-$0.78 \\
ES419 & 00:33:22.80 & $-$43:59:15.4 & 4.4 & $-$1.35 \\
ES427 & 00:34:11.59 & $-$43:58:17.0 & 22.3  & $-$1.08 \\
ES509 & 00:31:38.63 & $-$43:52:20.8 & 22.7  & $-$1.02 \\
ES749 & 00:29:05.23 & $-$43:34:03.9  & 10.3   & $-$1.08 \\
ES798 & 00:39:07.93 & $-$43:32:05.8  & 11.7   & $-$0.81 \\
ES973 & 00:38:44.14 & $-$43:19:20.4  & 11.6   & $-$1.15 \\
ES1259 & 00:38:27.171 & $-$42:51:33.8 &4.5\\
\\ 
CS114 & 03:27:59.89 & $-$27:55:54.7 & 7.7   & $-$1.34 \\
CS164 & 03:29:00.20   & $-$27:37:54.8 & 1.6  & $-$0.92 \\
CS215 & 03:29:50.02 & $-$27:31:52.6 	&1.6	& $-$0.76\\
CS241 & 03:30:10.22 & $-$28:26:53.0 & 1.0   & $-$1.96 \\
CS255 & 03:30:24.08 & $-$27:56:58.7 	& 0.5 \\
CS415 & 03:32:13.97 & $-$27:43:51.1 & 2.6 & $-$2.38 \\
CS538 & 03:33:30.20 & $-$28:35:11.2 & 2.0   & $-$0.88 \\
\hline\end{tabular}
\end{table*}

We selected 16 of the IFRS based on their flux density and spectral index.  The sources range in flux density from 0.5\,mJy to 23\,mJy and possess spectral indices from $-$2.4 to $-$0.2. An explanation of the calculation and derivation of these values, including the problems associated with resolution and scintillation, may be found in  Middelberg et al. (2011)\nocite{middelberg2010}, from which the IFRS values were taken. These parameters are certainly similar to those of the known pulsar population, which have flux densities that range from 0.01 to 1100\,mJy at an observing frequency close to 1400\,MHz and have known spectral indices ranging between $-3.5$ and $+0.9$ with a median of $-1.7$ \citep{mhth05}. It should be noted that the lower value of the pulsar flux density range is not due to any physical characteristic of pulsars, but is instead due to the physical limitations in the sensitivity of our instrumentation.

 In Table~\ref{tb:sources} we tabulate the source names, positions, flux density and spectral index of the IFRS.  For each source we carried out a 35 minute observation using the Parkes 64-m radio telescope equipped with the 20-cm 13-multibeam receiver (although only the central beam was utilised for our analysis), with 340\,MHz of bandwidth centred on 1352\,MHz. We used the Berkeley Parkes Swinburne Recorder (BPSR), a high resolution digital filterbank, providing 870 frequency channels, 2-bit samples every 64 $\mu$s. This observing setup is almost identical to the High Time Resolution Universe Pulsar Survey project \citep{kjv+10}, using a similar bandwidth, observing frequency backend system and sampling rate, with high sensitivity to millisecond pulsars. The data were recorded to magnetic tape for off-line processing.  The data were processed using the {\sc Hitrun} pipeline (see \citealp{kjv+10} for details) to search for periodic signals in dedispersed time series with trial dispersion measures (DMs) in the range of 0 to 1000\,cm$^{-3}$\,pc. 

We determine the theoretical sensitivity of our observations using the radiometer equation, which gives the fundamental limiting flux density of the central beam,
\begin{equation}
S_{min}=\frac{{\sigma}(T_{sys}+T_{sky})}{G\sqrt{2B{\tau}_{obs}}}\sqrt{\frac{W}{1-W}}.
\end{equation}
In this equation, $\sigma$ is the cutoff S/N for a positive detection, $T_{sys}$ is the system noise temperature, $T_{sky}$ is the sky noise temperature, $G$ is the system gain, $B$ is the observing bandwidth, $\tau_{obs}$ is the integration time of the observations and $W$ is the fractional pulse width. The typical value of $W$ for pulsars is approximately 10 percent of the pulse period, although it can be as high as 50 percent. 
%
Scaling our parameters from those of \citet{kjv+10}, setting $\sigma=8$, $T_{sys}=23$ K, $T_sky=0.7$ K and $G=0.735$, we arrive at a sensitivity to pulsed emission of $S_{min}=0.07$ mJy for a 10 percent pulse duty cycle, or $S_{min}=0.21$ mJy for a 50 percent pulse duty cycle. 
As all of our chosen IFRS candidates possess continuum flux densities above the higher of the two flux density limits, if any of them are indeed pulsars they should be easily detectable by our observations at the $8\sigma$ confidence level.

\section{Results and discussion}

The folded pulse profiles (intensity as a function of pulse phase) were viewed by eye for all candidates with a signal-to-noise ratio above $8\sigma$.
The observed pulse profiles for each candidate and their pulse periods were clearly caused by radio frequency interference or processing artefacts. For example, one candidate with a signal-to-noise ratio of $13 \sigma$ was detected with a frequency of 50.01 Hz and a sinusoidal pulse profile, indicating RFI originating with the mains power supply. We note that the minimum flux density of the chosen IFRS is 0.5 mJy, and the median flux density of the IFRS is approximately 4.4 mJy. If these flux densities are due to pulsars, then with a duty cycle of $W=0.5$ and an $8\sigma$ detection level of 0.21 mJy, the faintest of the IFRS should have been detected at a confidence level of $19\sigma$ with most of the sources ruled out by more than $100\sigma$. We conclude that none of the IFRS observed in this project show emission consistent with being a radio pulsar.

The absence of pulsar detections in this project emphasises the difficulty of identifying pulsar candidates through continuum surveys.  Despite the four detections described in Section \ref{sec:Intro}, the vast majority of continuum  (i.e., low time resolution) pulsar surveys have found no positive results (e.g. \citealp{kcc+00} and \citealp{kcb+00}). Also, while our analysis attempted to take advantage of the typically steep values of pulsar spectral indices in selecting our targets, the principle means of searching for pulsar candidates in continuum surveys has been to identify sources with significant linear and/or circular polarisation \citep{hmxq98, hdv+09}. Further studies using these selection criteria may yield better results.


Finally, the Evolutionary Map of the Universe (EMU) survey\footnote{http://www.atnf.csiro.au/people/rnorris/emu/}, to be carried out using the Australian Square Kilometre Array Pathfinder (ASKAP), is likely to detect 70 million radio sources with a sensitivity limit of $\sim 10 \mu$Jy. Over a million of these sources are likely to be currently undetected IFRS. However, at this point in time, we cannot make a prediction as to the number of these IFRS which may be pulsars, or indeed the fraction of pulsars expected in the total EMU source count. These values, as well as the techniques which may be used to more accurately isolate candidate pulsars from large-scale continuum surveys such as EMU, are the subject of ongoing research and are to be published in later work.

\section{Conclusions}

We have searched for short-term radio pulsations originating from 16 of the ATLAS infrared faint sources and find that pulsed emission cannot account for the observed radio flux density.  Since the sensitivity limit of our observations is well below the observed flux densities of the chosen IFRS, it is unlikely that any of these enigmatic sources are simply close-by pulsars in our own Galaxy. Hence, the nature of these sources is yet to be determined.

\section*{Acknowledgements} 

GH is the recipient of an Australian Research Council QEII Fellowship (project \#DP0878388).  The Parkes telescope is part of the Australia Telescope which is funded by the Commonwealth of Australia for operation as a National Facility managed by CSIRO.

\bibliography{journals,modrefs,psrrefs,crossrefs}
\bibliographystyle{mn2e}

\end{document}